\documentclass[12pt]{article}
\usepackage{jheppub}
\usepackage{amsmath}
\usepackage{epsfig,multicol}

\newcommand{\be}{ \begin{equation}}
\newcommand{\ee}{\end{equation}}
\newcommand{\bea}[1]{\begin{eqnarray}\label{#1} }
\newcommand{\eea}{\end{eqnarray}}

\def\ZZZ{{\hskip-3pt\hbox{ Z\kern-1.6mm Z}}}
\def\zzz{{\hskip-3pt\hbox{ z\kern-1mm z}}}

\newcommand{\half}{{1\over 2}}

\def\one{{\hbox{ 1\kern-.8mm l}}}
\def\zero{{\hbox{ 0\kern-1.5mm 0}}}

\title{Stringy Symmetries and the Higher Spin Square}

\author{Matthias R.\ Gaberdiel$^{a}$, and}
\author{Rajesh Gopakumar$^{b}$ }

\affiliation[a]{Institut f\"ur Theoretische Physik, ETH Zurich, \\
$\;$ CH-8093 Z\"urich, Switzerland }
\affiliation[b]{Harish-Chandra Research Institute, \\ Chhatnag Road, Jhusi,
Allahabad, India 211019}
\emailAdd{gaberdiel@itp.phys.ethz.ch}
\emailAdd{gopakumr@hri.res.in}

\abstract{Tensionless string theory on ${\rm AdS}_3\times {\rm S}^3\times \mathbb{T}^4$, 
as captured by a free symmetric product orbifold, has a large set of conserved currents which can be usefully 
organised in terms of representations of a ${\cal N}=(4,4)$ supersymmetric higher spin algebra. In this paper we focus 
on the single particle currents which generate the asymptotic stringy symmetry algebra on ${\rm AdS}_3$, and
whose wedge modes describe the unbroken gauge symmetries of string theory in this background. We show that this global
subalgebra contains two {\it distinct} higher spin algebras that generate  the full 
algebra as a `higher spin square'. The resulting unbroken stringy symmetry algebra is exponentially 
larger than the two individual higher spin algebras.}

\begin{document}

\maketitle

\section{Introduction}

The uniqueness of string (or M-)theory is one of its most attractive features, and yet the reason that  it is so, 
is not transparent. Indeed, until the advent of string dualities it was not even clear that the handful of  different 
looking perturbative string theories were limits of a single theory. Whenever we encounter in physics a rigidity of 
structure it is most often due to an underlying symmetry (or invariance) principle. Hence, we might try and look 
for such a symmetry principle constraining string theory. However, if a powerful constraining symmetry is present in 
string theory it is mostly hidden from view. Perturbative string theories expanded about flat space manifest 
diffeomorphism and gauge invariance in the massless sector, but little else. But there have been persistent 
hints \cite{Gross:1988ue, Witten:1988zd, Moore:1993qe, Sagnotti:2011qp}, e.g., in high energy scattering, 
of the restoration of a larger broken symmetry. If this is so, one needs a more symmetric vacuum to expand 
around, to manifest (at least some more of) the additional symmetries. 

String theories in ${\rm AdS}$ backgrounds might be better test cases for this purpose. 
Indirectly, through gauge-string duality, we are led to the conclusion that the ${\rm AdS}$ string theory 
dual to a free large $N$ QFT would have a large set of unbroken gauge symmetries 
\cite{Sundborg:2000wp, Witten, Mikhailov:2002bp, Sezgin:2002rt}. These are the symmetries of the 
Vasiliev higher spin theory \cite{Vasiliev:2003ev}. In fact, the Vasiliev theories of interacting higher spin 
gauge fields are a very elegant realisation of how invariance principles can largely constrain the form of a theory. 
At least when coupled to a finite number of low spin matter fields the classical equations of motion appear to be 
unique (perhaps up to some parameters).\footnote{The classical and 1-loop Vasiliev theory successfully 
capture the large $N$ limit of vector-like models with no propagating gauge fields 
\cite{Klebanov:2002ja, Sezgin:2003pt, Giombi:2009wh, Giombi:2010vg, Gaberdiel:2010pz, %
Aharony:2011jz, Giombi:2011kc, Giombi:2013fka, Giombi:2014iua}, see \cite{Gaberdiel:2012uj, Giombi:2012ms} 
for reviews.} But string theory is not simply a Vasiliev theory --- it has an infinite set of matter fields (of 
arbitrarily high spin) coupled to the Vasiliev gauge fields. Also the gauge symmetries of the Vasiliev theory are 
only one infinite tower or one Regge trajectory of the string theory. So this is not quite a large enough unbroken 
stringy symmetry --- a genuinely stringy symmetry should be exponentially larger in size. 
Are there vacua which exhibit such a larger unbroken symmetry? 

String theories on ${\rm AdS}_3$ have an even larger unbroken symmetry than a generic 
${\rm AdS}_{d+1}$ background. Again this can be indirectly inferred from gauge-string duality. For free gauge 
theories in $d\geq 3$, the only single trace conserved currents are built from bilinears and their derivatives, 
as follows from unitarity bounds for conserved currents in a CFT. In the bulk these correspond exactly to the 
Vasiliev gauge symmetries and no more. But CFTs in two dimensions which are dual to ${\rm AdS}_3$ 
string backgrounds are, in many cases, believed to be deformations of free symmetric product orbifolds, 
see \cite{David:2002wn} for a review. These have a very large set of conserved currents at their 
orbifold point --- those built from bilinears of the underlying free bosons or fermions form only a tiny subset. 
As we will see below, this is true even after isolating the single particle (`single trace') generators from 
amongst all the currents.  Assuming that this free point corresponds to a tensionless point in the moduli 
space of ${\rm AdS}_3$ string vacua, we conclude that there must be a much larger unbroken gauge symmetry 
in the bulk string theory here than in the higher dimensional cases \cite{Gaberdiel:2014cha}. 

We should, however, keep in mind that string theory (or any theory of gravity, for that matter) on ${\rm AdS}_3$ 
always exhibits an enhanced asymptotic symmetry algebra. Indeed, as first observed by Brown and Henneaux, 
diffeomorphism invariance on ${\rm AdS}_3$ has its global algebra of 
$\mathfrak{sl}(2, \mathbb{R})\oplus \mathfrak{sl}(2, \mathbb{R})$ enhanced to two copies of the Virasoro algebra 
with a universal central charge \cite{Brown:1986nw}. 
Similarly, the higher spin gauge invariances of a Vasiliev theory 
have their global symmetry algebra of $\mathfrak{sl}(N) \oplus \mathfrak{sl}(N)$ (or more generally, 
${\rm hs}[\lambda]\oplus {\rm hs}[\lambda]$) enhanced to two copies of a ${\cal W}_N$
(or ${\cal W}_{\infty}[\lambda]$, respectively) asymptotic symmetry algebra 
\cite{Henneaux:2010xg, Campoleoni:2010zq, Gaberdiel:2011wb, Campoleoni:2011hg}. 
In order to extract the global part of the asymptotic symmetry algebra of string theory 
we have to restrict it to its `wedge' modes --- these consist 
of the `single particle'
generators that annihilate both the in- and out-vacuum --- and take the central charge to infinity. 
The resulting `wedge' algebra --- for example, for the case of the bosonic ${\cal W}_N$ algebra 
this is just $\mathfrak{sl}(N)$ --- 
will then tell us about the unbroken stringy symmetries in the bulk. It is also this
algebra which is gauged and which will, hopefully,
help in constraining the structure of string theory ---  just as the Vasiliev theory based 
on ${\rm hs}[\lambda]\times {\rm hs}[\lambda]$ was largely so, by its algebra.
\smallskip

In this paper we make a beginning towards getting an explicit handle on this unbroken stringy algebra for the 
case of superstrings on ${\rm AdS}_3\times {\rm S}^3\times \mathbb{T}^4$ in its tensionless limit, as captured 
by the symmetric product orbifold SCFT of $\mathbb{T}^4$. We will exploit the fact that the whole symmetric orbifold chiral
algebra can be 
organised in terms of representations of the superconformal ${\cal W}_\infty^{({\cal N}=4)}$ 
higher spin algebra  \cite{Gaberdiel:2014cha};
the relevant representations are of the form $(0;\Lambda)$, where 
$\Lambda$ is a representation of ${\rm SU}(N)$ \cite{Gaberdiel:2013vva, Gaberdiel:2014cha}. 
First, we will identify which of these representations correspond to {\it single particle} symmetry generators. 
These are the analogues of `single trace' operators in gauge theories, and are built from a single cycle of the 
symmetric group. Their wedge modes then generate the global part of the symmetry algebra which, 
as mentioned above, is the unbroken gauge algebra of the bulk string theory. 

For the case of the symmetric orbifold of $\mathbb{T}^4$, the single particle generators are 
in one-to-one correspondence with the elements of the chiral algebra of $\mathbb{T}^4$. (Indeed, to each 
element $\rho$ in the chiral algebra of $\mathbb{T}^4$, we can associate a generator of the symmetric 
orbifold algebra via the symmetrised sum, $\sum_i \rho^i$.) Note that when we take the large $N$ 
limit of the symmetric orbifold, all these generators become independent (unlike the 
generators of the chiral algebra of a single $\mathbb{T}^4$ which cannot all be independent). 
The results of  \cite{Gaberdiel:2014cha} imply that these resulting generators must organise themselves
in terms of representations of the wedge subalgebra ${\rm shs}_2[\lambda=0]$ of 
${\cal W}_{\infty}^{({\cal N}=4)}[0]$ --- the representation of the full ${\cal W}_\infty^{({\cal N}=4)}[0]$
algebra will also contain multi-particle states, and it is only the wedge modes that map single particle
generators to themselves. We find that the relevant representations (which carry the same label 
both for ${\cal W}_{\infty}^{({\cal N}=4)}$ and its wedge subalgebra), are $(0;\Lambda)$,  where $\Lambda$
is described in terms of Dynkin labels as $\Lambda=[m, 0, \ldots, 0,n]$, and each representation appears with 
multiplicity one. In simpler terms, these are the 
completely symmetric powers of the fundamental/antifundamental representation.  
Here the generators from ($m=1$, $n=0$), ($m=0$, $n=1$) and ($m=n=1$) 
are those of the ${\cal W}_{\infty}^{({\cal N}=4)}$ 
algebra itself (the free fermions, and the invariant bilinears together with their superconformal descendants). 
The others account for the additional single particle generators of the full symmetric orbifold chiral
algebra.

In order to describe the Lie algebra structure of these wedge generators, it is 
convenient to go back to the full asymptotic symmetry
algebra. Furthermore, it is instructive to analyse first the structure of a simpler
bosonic toy model, the $N$'th symmetric orbifold of a single real boson. In this case the single particle 
generators of the chiral algebra are in one-to-one correspondence with the chiral sector of a single boson. While this 
provides a complete characterisation of  the algebra, it is useful to refine this description in terms of its 
higher spin subalgebras. This is what the `higher spin square' structure uncovers. 

A natural ${\cal W}_\infty$ subalgebra, in this case, is spanned by the bilinear expressions
\be\label{1.1}
\sum_{i=1}^{N}\,  (\partial^{m_1} \phi^i)\,  (\partial^{m_2} \phi^i) \ , \qquad m_1,m_2\geq 1 \ , 
\ee
suitable linear combinations of which define the quasiprimary generators of spin \linebreak $s=m_1+m_2$. They
generate the even spin subalgebra ${\cal W}^{(\rm{e})}_\infty[1]$ of ${\cal W}_\infty[1]$. On the other hand, the 
most general `single-particle' generators involve the symmetrised products of order $p\geq 1$
\be\label{ppower}
\sum_{i=1}^{N} \, (\partial^{m_1} \phi^i) \cdots (\partial^{m_p} \phi^i ) \ , \qquad 
m_1,\ldots, m_p\geq 1 \ .
\ee
The (singular part of the) OPE of a ${\cal W}_\infty$ generator with an order $p$ generator (\ref{ppower}) contains,
up to possible terms of lower order, only generators of order $p$; corrected by suitable lower order terms,
the generators of order $p$ therefore form a representation of ${\cal W}_\infty$.\footnote{Strictly speaking, 
they only form a representation of the wedge subalgebra of ${\cal W}_\infty$. If we want
to think of them as honest representations of ${\cal W}_\infty$ we must also include the multiparticle
states that are made up of one generator of order $p$, together with an arbitrary
number of bilinear ${\cal W}_\infty$ generators.}
In order to visualise this, 
we can arrange these representations in parallel vertical columns, where each column comprises 
of `descendants' with the ${\cal W}^{(\rm{e})}_\infty[1]$ algebra acting `vertically', see Figure~1. In other words, the 
commutators with the generators from the higher spin algebra 
${\cal W}^{(\rm{e})}_\infty[1]= {\cal W}^{({\rm vert})}_\infty$ move one up and down along each of the green columns. 
Note that each column really becomes `thicker' as we move vertically downwards, corresponding to the 
exponentially increasing number of descendants, i.e., of splitting up the total spin among the $p$ integers
$m_1,\ldots,m_p$. These columns generalise to the representations $(0;\Lambda)$ with 
$\Lambda=[m, 0, \ldots, 0,n]$ in the ${\cal N}=4$ case. 
\smallskip

The crucial observation is now that the commutators involving the `top' generators from each
column {\it also} define a higher spin symmetry algebra, the `horizontal' ${\cal W}$-algebra
${\cal W}_{\infty}^{({\rm hor})}$. In fact, the generators associated to 
$\sum_i(\partial \phi^i)^l$ close, up to lower order correction terms, to form the 
${\cal W}_\infty$ algebra ${\cal W}_{1+\infty}[0]$; this was first observed in \cite{Pope:1991ig},
and is basically a consequence of the fact that we may equivalently describe the real boson theory in terms
of a complex fermion, whose associated ${\cal W}_\infty$ algebra is ${\cal W}_{1+\infty}[0]$.
We can therefore equally well organise the full stringy algebra in terms of representations of this
horizontal ${\cal W}$-algebra (or rather again its wedge subalgebra), i.e., in terms of the red rows 
of Figure~1, see eq.~(\ref{3.21}) below. From this perspective, the left-most terms in each row
(except for the first row) are the bilinear currents of the form (\ref{1.1}) with spin $s\geq 3$, while the 
top left entry corresponds to $\sum_i \partial^l \phi^i$, i.e., to 
the spin $s=1$ generator of ${\cal W}_{1+\infty}[0]$, together with its $L_{-1}$ descendants --- these
states sit in a single irreducible representation of the (wedge subalgebra of the) 
vertical algebra ${\cal W}^{({\rm vert})}_\infty$.

\begin{figure}[t]
\begin{center}
\includegraphics[width=15.5cm]{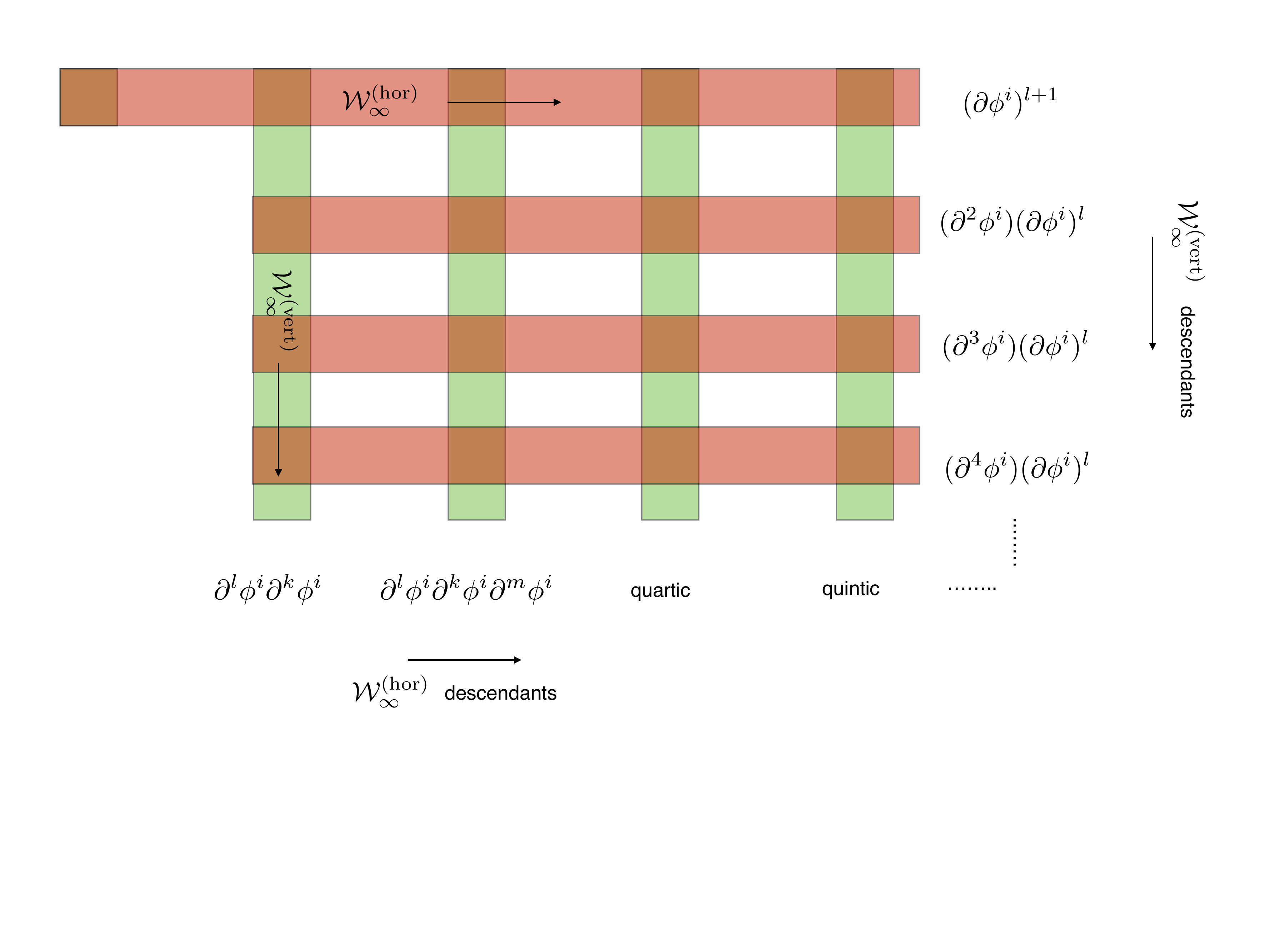}
\end{center}
\vspace*{-2.5cm}
\caption{The Higher Spin Square.}
\end{figure}

The picture we thus arrive at is that the commutators along a given row are governed by the horizontal higher spin algebra 
${\cal W}_{\infty}^{({\rm hor})}$, while commutators down a given column are governed by the vertical 
higher spin algebra ${\cal W}_{\infty}^{({\rm vert})}$. Knowing both thus determines, in 
principle, an arbitrary commutator between elements of the chiral algebra. 
We may use, for instance, ${\cal W}_{\infty}^{({\rm vert})}$ to bring any element to the 
topmost row of its column, and then employ ${\cal W}_{\infty}^{({\rm hor})}$ to evaluate commutators
with any other entry.\footnote{Or we may just as well employ ${\cal W}_{\infty}^{({\rm hor})}$ to bring any 
element to the left most column, and then utilise ${\cal W}_{\infty}^{({\rm vert})}$.} 
Therefore, the global horizontal and vertical higher spin symmetry algebras together effectively 
generate the full stringy symmetry --- the wedge algebra of the above chiral algebra. We will thus refer to 
this symmetry structure as a `higher spin square'.\footnote{The terminology must not 
mislead one into viewing this as a square (or tensor product) of the two higher spin symmetries. 
The rank is, in a precise sense, exponentially larger than that of either the vertical or horizontal symmetry 
algebra since it involves generators built effectively from multilinear products of the individual higher spin 
generators.} We believe the existence of this higher spin square is 
an indication that the stringy symmetries are best understood by decomposing them in terms of the higher 
spin symmetries. 
\smallskip

While the analysis for this bosonic toy model is rather natural and symmetrical (and can be checked fairly
directly, see in particular Section~4.1), we have so far not succeeded in finding 
an equally canonical description in the ${\cal N}=4$ case, although a number of possible constructions 
exist (see Section~4.3). On the other hand, we have found a 
neat finite-dimensional analogue --- associated to the Clifford
algebra --- which builds intuition for this structure, and that is studied in some detail in Section~4.2.
In that case, one has $\gamma$ matrices in $2n$ dimensions. 
The bilinears in the $\gamma$ matrices are the generators of $\mathfrak{so}(2n)$ which plays the role of the 
vertical algebra. The other multi-linears define various antisymmetric representations of $\mathfrak{so}(2n)$ 
and can be arranged in vertical columns.  The $\gamma$-matrix algebra gives us a natural set of commutators 
for elements across different columns --- a horizontal algebra --- and the full Clifford algebra is generated by the 
horizontal and vertical algebras. In this case it is easy to see that the resulting Lie algebra --- the `Clifford 
algebra square'  ---  is just $\mathfrak{su}(2^n)$ since the gamma matrices generate all linearly independent 
matrices of size $2^n$. 
This shows that there is an exponential enhancement in rank in going from the vertical algebra 
$\mathfrak{so}(2n)$ to the full Clifford algebra $\mathfrak{su}(2^n)$. We find this to be a useful model to keep in mind as some kind of a finite 
truncation of the higher spin square. 
\medskip

The outline of this paper is as follows. Section~2 contains the identification of the single particle symmetry generators
for the symmetric orbifold of $\mathbb{T}^4$. 
Section~3 discusses the decomposition of the chiral algebra for free bosons and free fermions in terms of both horizontal and vertical bosonic 
${\cal W}_{\infty}$ representations. These building blocks are then put 
together for the SUSY ${\cal N}=4$ case. 
Section~4 uses these results and describes the higher spin square, first in the case of the free boson (Section~4.1) and then 
for the ${\cal N}=4$ case (Section~4.3). We also  introduce the Clifford algebra square in Section~4.2.
Finally Section~5 contains miscellaneous closing remarks.

\section{Single Particle Symmetry Generators}\label{sec:symmetric}

Let us begin by identifying the single particle currents of the symmetric orbifold of $\mathbb{T}^4$.
The basic idea is to write the chiral sector of the orbifold partition function 
for ${\rm Sym}^N(\mathbb{T}^4)$ (in the limit of large $N$) in a second quantised form. 
Since the operation of taking the symmetric product is essentially one of `multi-particling', it is not greatly 
surprising that the generating function of the single particle currents is the same as that of four free bosons 
and fermions --- the content of the supersymmetric $\mathbb{T}^4$ theory. The new observation will be the 
decomposition of these generators into specific representations of the small ${\cal N}=4$  
${\rm shs}_2[\lambda=0]$ algebra, the wedge subalgebra of the supersymmetric ${\cal W}_\infty$ algebra.

Let us start with the familiar formula for the generating function for the untwisted sector of a general  
symmetric orbifold \cite{Dijkgraaf:1996xw}
\be\label{generating}
\sum_{N=0}^{\infty} p^N Z_{\rm RR}^{({\rm U})} ({\rm Sym}^k(X)) = \prod_{\Delta,\bar\Delta,\ell,\bar\ell} 
\frac{1}{(1 - \, p q^{\Delta} \bar{q}^{\bar\Delta} y^{\ell} \bar{y}^{\bar{\ell}} )^{ 
c(\Delta,\bar\Delta, \ell,\bar\ell) }} \ , 
\ee
where the coefficients $c(\Delta,\bar\Delta,\ell,\bar\ell)$ are the expansion coefficients of the
R-R partition function of $X$ (with the insertion of $(-1)^{F+\tilde{F}}$),
\be
Z_{\rm RR}(X) = \sum_{\Delta,\bar\Delta,\ell,\bar\ell}   c(\Delta,\bar\Delta, \ell,\bar\ell) \, 
q^{\Delta} \bar{q}^{\bar\Delta} y^{\ell} \bar{y}^{\bar{\ell}} \ .
\ee
We are interested in the ${\cal W}$ algebra of the symmetric orbifold of $X=\mathbb{T}^4$. 
In particular, we want to analyse only 
the purely left-moving states and we want to describe them in the NS-sector. Explicitly applying the 
spectral flow, this leads to 
\be\label{genNS}
\sum_{N=0}^{\infty} p^N\, Z^{(N)}_{\rm vac, NS}(\tau,y)  = \prod_{n=0}^{\infty} \prod_{\ell \in \mathbb{Z}} 
\frac{1}{\bigl( 1 + (-1)^\ell \, p\, q^{n+\ell/2+1/4} \, y^{\ell+1}\bigr)^{c(n,\ell)}} \ , 
\ee
where the $c(n,\ell)$ are the expansion coefficients of the R-sector chiral partition function
of $\mathbb{T}^4$ (with the insertion of $(-1)^F$)
\be\label{cdef}
(y - 2 + y^{-1}) \prod_{n=1}^{\infty} \frac{ (1 - y q^n)^2 (1-y^{-1} q^n)^2}{(1-q^n)^4}  = 
\sum_{n=0}^{\infty} \sum_{\ell \in\mathbb{Z}} c(n,\ell) q^n y^\ell \ . 
\ee
Note that the term with $n=0$ and $\ell=-1$ in eq.~(\ref{genNS}) describes the contribution of the NS vacuum
with $q^{-1/4}$ --- it follows from eq.~(\ref{cdef}) that $c(0,\pm 1)=1$ --- while all other
terms in eq.~(\ref{genNS}) lead to positive powers of $q$. Thus 
the NS vaccuum character stabilises for sufficiently large $N$ to 
\be
q^{N/4} \, Z^{(N)}_{\rm vac, NS}(\tau,y)  = \prod_{n=0}^{\infty} \prod_{\ell \in \mathbb{Z}} {}^\prime
\frac{1}{\bigl( 1 + (-1)^\ell \, q^{n+\ell/2+1/2} \, y^{\ell+1}\bigr)^{c(n,\ell)}} \ , 
\ee
where the prime at the product means that now the term with $n=0$ and $\ell=-1$ is excluded. 
This can now also be written directly in terms of the NS sector expansion coefficients 
\be\label{NSex}
q^{N/4} \, Z^{(N)}_{\rm vac, NS}(\tau,y)  = \prod_{r=\frac{1}{2}}^{\infty} \prod_{\ell \in \mathbb{Z}} 
\frac{1}{\bigl( 1 - (-1)^{2r}  \, q^{r} \, y^{\ell}\bigr)^{d(r,\ell)}} \ , 
\ee
where $r$ runs over half-integers and integers and the coefficients $d(r,\ell)$ are now the expansion
coefficients of the spectrally flowed version of (\ref{cdef})
\be\label{NSpart}
\prod_{n=1}^{\infty} \frac{(1 - y q^{n-\frac{1}{2}})^2 (1 - y^{-1} q^{n-\frac{1}{2}})^2}{(1-q^n)^4} = 1 + 
\sum_{r=\frac{1}{2}}^{\infty} \sum_{\ell\in\mathbb{Z}} d(r,\ell)\, q^r  y^\ell \ . 
\ee


The expression in eq.~(\ref{NSex}) is now very suggestive: it is already of `multi-particle' form, where each 
state appearing in the NS character of $\mathbb{T}^4$ in eq.~(\ref{NSpart}) defines a separate particle. 
Note that because of the $L_{-1}$ descendants, each holomorphic current 
of spin $s$ contributes not just as 
$1/(1-q^s)$ (if bosonic) or $(1+q^s)$ (if fermionic), but rather as 
\be\label{wedgeout}
\prod_{n=0}^\infty \frac{1}{(1-q^{s+n})} \quad \hbox{(bosonic)} \qquad \hbox{or}\qquad 
\prod_{n=0}^\infty (1+q^{s+n}) \quad \hbox{(fermionic)} \ .
\ee
Thus the full chiral algebra of the symmetric orbifold is generated by $N(r,\ell)$ fields
of spin $r$ and $\mathfrak{u}(1)$ charge $\ell$ where
\be\label{Ndef}
(1-q) \, \Bigl[ 
\prod_{n=1}^{\infty} \frac{(1 + y q^{n-\frac{1}{2}})^2 (1 + y^{-1} q^{n-\frac{1}{2}})^2}{(1-q^n)^4}  -1 \Bigr]
= \sum_{r=\frac{1}{2}}^{\infty} \sum_{\ell \in \mathbb{Z}} N(r,\ell) \, q^r\, y^\ell \ . 
\ee
Equivalently, we can then write the character of the chiral algebra of the stringy extension as 
\be\label{NSex1}
q^{N/4} \, Z^{(N)}_{\rm vac, NS}(\tau,y)  = \prod_{r=\frac{1}{2}}^{\infty} \prod_{\ell \in \mathbb{Z}} 
\, \prod_{n=0}^{\infty} \frac{1}{\bigl( 1 - (-1)^{2r}  \, q^{r+n} \, y^{\ell}\bigr)^{ (-1)^{2r}N(r,\ell)}} \ .
\ee

As mentioned earlier, this result for the single particle generators {\it per se} 
may not come as a big surprise since  
the process of taking the symmetric product is essentially one of `multi-particling'. 
Thus it is 
natural that the single particle generators of the $\mathbb{T}^4$ (or, more accurately, $\mathbb{R}^4$) 
theory are in one-to-one correspondence with the chiral generators of four free bosons and their four 
fermionic superpartners (as captured in the generating function (\ref{Ndef})).\footnote{Note that we can 
only make a correspondence and not an identification 
since each independent generator is a fully symmetrised expression involving (derivatives of) all $4N$ bosons and fermions.}
What is important from our present point of view is to organise these symmetry generators in terms of 
the representations $(\Lambda_+; \Lambda_-)$ of the wedge algebra of the 
super ${\cal W}_{\infty}^{(4)}[0]$ algebra. There is, in fact, 
a simple way to characterise the representations which contribute to the single particle generators: they 
correspond to 
\be\label{wreps}
(\Lambda_+; \Lambda_-) = \bigl(0;[m,0,\ldots.0,n] \bigr) \ .
\ee
This is the natural class of representations to consider since they are the only ones which contain exactly a
single cycle singlet of the symmetric group (see the discussion in appendix~C.1 of \cite{Gaberdiel:2014cha}).
This single cycle singlet is the analogue of the single trace operators when one 
gauges the symmetric group. 

Combining this observation with the earlier generating function 
then leads to the prediction for an identity for the wedge 
characters\footnote{The full characters would also include multi-particle terms, and thus only the wedge 
characters appear here.}
of these super ${\cal W}_{\infty}^{(4)}[0]$ representations
\begin{eqnarray}\label{wedgegen}
\sum_{r,\, \ell} N(r, \ell) \, q^r\, y^\ell & = & (y+y^{-1}) \Bigl[ 2 q^{1/2} + 4 \sum_{r=\frac{3}{2},\frac{5}{2},\ldots} q^r  \Bigr]
+ \sum_{n=1}^{\infty} \, (y^2 + 6 + y^{-2}) q^n \qquad \label{wcont} \\
& & {} + (1-q)\, \sum_{m,n\geq 0}{}^{'} \chi^{({\rm wedge}) \, {\cal N}=4[0]}_{(0;[m,0,\ldots,0,n])} (q,y) \ . \label{wprim}
\end{eqnarray}
Here the terms on the RHS in (\ref{wcont}) describe the ${\cal W}_{\infty}^{(4)}[0]$ generators, while the contributions from
the sum in (\ref{wprim}) account for the wedge modes of the single-particle representations (\ref{wreps}),
with the terms generated from $L_{-1}$ subtracted out.\footnote{The prime in the sum indicates that the cases 
$(m,n)=(1,0), (0,1), (1,1)$ are excluded. In effect, these terms precisely describe the contribution of 
the first line, i.e., of (\ref{wcont}).}
Note that we have to restrict to the `wedge' part of the characters, since the other contributions
correspond to `multi-particle' states involving, in addition to a generator from $\bigl(0;[m,0,\ldots.0,n] \bigr)$, an arbitrary number of ${\cal W}_{\infty}^{(4)}[0]$ fields.  

We have tested this identity, using eq.~(\ref{Ndef}) as well as the wedge characters from 
\cite{Gaberdiel:2014cha},  up to order $q^{6}$. In the next section we shall also see 
how to understand it in terms of bosonic and fermionic building blocks.

\section{Building Blocks}

The decomposition of the modes of four free bosons and four free fermions into characters of the
wedge algebra of  ${\cal W}_{\infty}^{(4)}[0]$  is a generalisation of simpler decompositions of the chiral sector  
of a single free boson or fermion into characters of the wedge subalgebras of the corresponding bosonic 
${\cal W}_{\infty}$ algebras, as already sketched in the introduction. 
In this section we first exhibit these simpler decompositions and then explain an identity involving 
two distinct ${\cal W}_{\infty}$ algebras, see eq.~(\ref{3.21}), that is at the heart of the two dual 
higher spin square decompositions. Combining these simpler decompositions will also 
allow us to prove eq.~(\ref{wedgegen}), see the discussion in  Section~\ref{sec:together}.

\subsection{Free Boson}

To understand the identity in its simplest incarnation, consider the 
chiral algebra generated by the symmetric orbifold of a single free real boson. The analogue of 
eq.~(\ref{wedgegen}), in conjunction with eq.~(\ref{Ndef}), is in this case 
\be\label{bosid}
(1-q)\prod_{n=1}^{\infty}\frac{1}{(1-q^n)} =\sum_r N(r) \, q^r =
 (1-q)\Bigl(1+ \sum_{n=1}^{\infty}b^{({\rm wedge}) [\lambda=1]}_{([0^{n-1},1,0,\ldots,0];0)} (q) \Bigr) \ .
\ee
Here the first term is, up to the $(1-q)$ prefactor, the partition function of a single free boson 
while the last is the sum of the 
bosonic wedge characters
$b^{({\rm wedge})}_{(\Lambda;0)} (q)$ of the (even) ${\cal W}_{\infty}^{({\rm e})}[1]$ algebra
of \cite{Gaberdiel:2011nt}, see also \cite{CGKV}. Note that the ${\cal W}$-algebra
generators themselves are counted by the wedge character corresponding to $n=2$, see
eq.~(\ref{boswed}) below, 
\be
b^{({\rm wedge})[\lambda=1]}_{([0,1,0,\ldots,0];0)} (q) = \frac{q^2}{(1-q) (1-q^2)}  \ . 
\ee
Indeed, taking away the $(1-q)$ factor which just accounts for derivatives, this leads to a spectrum
of a single higher spin field for each even spin (which is what the theory of real bosons gives rise to).

In this case we can write down the branching functions (and therefore, wedge characters) explicitly following the results derived in 
\cite{Gaberdiel:2011zw}. The wedge characters are 
given by Schur polynomials (see eq.~(2.8) and below of \cite{Gaberdiel:2011zw}), which are 
essentially ${\rm U}(\infty)$ characters. Simplified expressions for these have been given in Appendix~A 
of \cite{Gaberdiel:2011zw},  see eqs.~(A.13), (A.14) and (A.16). In particular, 
\be
\chi^{{\rm U}(\infty)}_{[n,0,\ldots,0]}(q) = \frac{q^{\frac{n}{2}}}{\prod_{k=1}^n(1-q^k)}  \ ,
\ee
using the fact that 
\be
\prod_{(ij)\in [n,0,\ldots,0]}(1-q^{h_{ij}}) = \prod_{k=1}^n(1-q^k) \ ,
\ee
where $h_{ij}$ are the hook lengths of the boxes in the Young tableaux. 
After multiplying by the factor of $q^{\frac{\lambda}{2} n}$ (see eq.~(2.8) of \cite{Gaberdiel:2011zw}) 
and using the usual transposition rule between ${\rm U}(\infty)$ and coset Young tableaux, 
we have (for $\lambda=1$),
\be\label{boswed}
b^{({\rm wedge})[\lambda=1]}_{([0^{n-1},1,0,\ldots,0];0)} (q) = \frac{q^{n}}{\prod_{k=1}^n(1-q^k)} \ .
\ee
Thus the identity of eq.~(\ref{bosid}) reduces to showing 
\be\label{bosid1}
\prod_{n=1}^{\infty}\frac{1}{(1-q^n)} = 1+ \sum_{n=1}^{\infty}\frac{q^{n}}{\prod_{k=1}^n (1-q^k)} \ .
\ee
This identity is in fact a particular case of the so-called $q$-binomial theorem (see for example Section~10.2 of \cite{AAR}). 
Actually, in our case we can easily prove this identity using the combinatorics of a free scalar field which also sheds 
light on the nature of the associated chiral algebra generators. 
We organise the Fock space of the left moving excitations of a bosonic field as the union of $n$-particle excitations
\be\label{bosmodes}
\alpha_{-m_1}\alpha_{-m_2}\cdots\alpha_{-m_n}|0 \rangle = \alpha_{-1-k_1}\alpha_{-1-k_1-k_2}\cdots
\alpha_{-1-\sum_i^nk_i} |0\rangle \ .
\ee
On the right-hand-side we have written out the modes $m_i\geq 1$ in ascending order so that $k_i \geq 0$. 
The different contributions of such terms to the partition function is precisely given by 
$\frac{q^{n}}{\prod_{k=1}^n (1-q^k)}$. Thus summing over $n$ gives us indeed eq.~(\ref{bosid1}).
\smallskip

For the following it will also be important to consider the case of a complex boson, i.e., two 
real bosons. Using the fact that the wedge characters associated to boxes and anti-boxes 
simply mutliply, this can then be written as 
\be\label{complxbos}
\prod_{n=1}^{\infty} \frac{1}{(1-q^n)^2} = \sum_{n,m=0}^{\infty} 
b^{({\rm wedge})[\lambda=1]}_{([0^{n-1},1,0,\ldots,0,1,0^{m-1}];0)}(q) \ , 
\ee
where we have used that the term with $m=n=0$ is just the identity. In this case, the ${\cal W}$-algebra
generators themselves can be obtained from the wedge character with $m=n=1$,
\be
b^{({\rm wedge})[\lambda=1]}_{([1,0,\ldots,0,1];0)}(q) = \frac{q^2}{(1-q)^2} \ , 
\ee
accounting indeed for one field of each integer spin $s\geq 2$; this then generates ${\cal W}_\infty[1]$, without
any truncation to even spin.

\subsection{Free Fermion}\label{sec:ff}

There is also a fermionic version of the above identity, for which the relevant wedge characters are 
evaluated at $\lambda=0$. It takes the form\footnote{\label{foot5} In the free fermion case, we have the equivalence
of representations $(\Lambda;0) \cong (0;\Lambda^\ast)$ \cite{Gaberdiel:2011aa}. Thus, at this stage, it is
a matter of convenience to write this expression in terms of the wedge characters of $(0;\Lambda)$, rather
than $(\Lambda;0)$.}
\be\label{fersid}
(1-q)\prod_{r=1}^{\infty} (1+y \, q^{r-1/2}) =\sum_r \tilde{N}(r,\ell) \, q^r\, y^{\ell} =
(1-q)\Bigl(1+ \sum_{n=1}^{\infty}b^{({\rm wedge})[\lambda=0]}_{(0;[n,0,\ldots,0])} (q,y) \Bigr) \ .
\ee
At $\lambda=0$ the relevant wedge characters equal 
\be\label{ferwed}
b^{({\rm wedge})[\lambda=0]}_{(0;[n,0,\ldots,0])} (q,y) = \frac{y^n q^{n^2/2}}{\prod_{k=1}^n(1- q^k)} \ ,
\ee
and again the term with $n=2$ corresponds to the ${\cal W}_\infty^{(\rm e)}[0]$ generators themselves,
\be
b^{({\rm wedge})[\lambda=0]}_{(0;[2,0,\ldots,0])} (q) = \frac{y^2 q^{2}}{(1- q)(1- q^2)} \ .
\ee
The above identity is then a direct consequence of another special instance of a $q$-binomial theorem,
\be\label{fermid}
\sum_{k=0}^{\infty}  \frac{y^k q^{k^2/2}}{\prod_{l=1}^k(1-q^l)} =\prod_{r=0}^{\infty}(1+y \, q^{r+1/2}) \ . 
\ee
Once again, we can easily prove this identity using the combinatorics of a free fermion field.
As before, we organise the Fock space of the left moving excitations of a fermionic field as the union of
$n$-particle excitations
\be
\psi_{-m_1-\half}\psi_{-m_2-\half}\cdots\psi_{-m_n-\half}|0 \rangle = \psi_{-\half-k_1}\psi_{-\frac{3}{2}-k_1-k_2}\cdots
\psi_{-\frac{2n-1}{2}-\sum_i^nk_i} |0\rangle \ .
\ee
On the right-hand-side we have written out the modes $m_i\geq 0$ in ascending order so that $k_i \geq 0$. 
The different contributions of such terms to the partition function is precisely given by 
$$
\frac{y^n q^{\frac{n^2}{2}}}{\prod_{k=1}^n (1-q^k)}\ . 
$$
Thus summing over $n$ gives us indeed eq.~(\ref{fermid}).

\subsection{Bosonisation}

Actually, one can combine both fermionic and bosonic versions of these identities into dual
identities for a single generating function. The basic idea is that, because of bosonisation, the
neutral sector of the theory of a complex fermion is equivalent to that of a real boson.
For a complex fermion, the relevant decomposition identity equals
\be\label{fersid2}
\prod_{r=1}^{\infty} (1+y\, q^{r-1/2})\, (1+y^{-1} q^{r-1/2})  
= \sum_{n,m=0}^{\infty}b^{({\rm wedge})[\lambda=0]}_{(0;[n,0,\ldots,m])} (q,y)  \ .
\ee
Here we have used that the conjugate of the wedge characters (\ref{ferwed}) take the form 
\be\label{ferwedc}
b^{({\rm wedge})[\lambda=0]}_{(0;[0,\ldots,0,n])} (q,y) = \frac{y^{-n} \,q^{n^2/2}}{\prod_{k=1}^n(1- q^k)} \ ,
\ee
as well as the fact that the wedge characters associated
to boxes and anti-boxes multiply. Then noting that the $1$ in eq.~(\ref{fersid}) just corresponds to $n=0$, 
the identity eq.~(\ref{fersid2})
follows directly from the square of (\ref{fersid}). Note that in this 
language, the associated ${\cal W}_\infty$ algebra now corresponds to the sector $(0;[1,0,\ldots,0,1])$
for which the wedge character equals
\be
b^{({\rm wedge})[\lambda=0]}_{(0;[1,0,\ldots,0,1])} (q) = \frac{q^{1}}{(1-q)^2} \ .
\ee
This then leads to one generator for each integer spin (including $1$), i.e., to ${\cal W}_{1+\infty}[0]$. 

\noindent If we restrict the complex fermion theory to the neutral sector, we have the identity
\be
\left. \prod_{r=1}^{\infty} (1+y q^{r-1/2}) (1+y^{-1} q^{r-1/2}) \right|_{y^0} = \prod_{n=1}^{\infty} \frac{1}{(1-q^n)} \ , 
\ee
i.e., the neutral sector of two complex fermions agrees precisely with the theory of a single real boson. Thus 
we can rewrite the bosonic partition function also in terms of the `neutral' wedge characters of the $\lambda=0$
theory, 
\be\label{bosid2}
(1-q)\prod_{n=1}^{\infty}\frac{1}{(1-q^n)} =\sum_r N(r) \, q^r =
 (1-q)\Bigl(1+ \sum_{n=1}^{\infty}b^{({\rm wedge}) [\lambda=0]}_{(0;[n,0,\ldots,0,n])} (q) \Bigr) \ .
\ee
This is equivalent to the identity
\be\label{newid}
\prod_{n=1}^{\infty} \frac{1}{(1-q^n)} = 1 + \sum_{n=1}^{\infty} \frac{q^{n^2}}{\prod_{l=1}^{n} (1-q^l)^2} \ ,
\ee
which is indeed true (see Corollary 10.9.4 and below of \cite{AAR}). Thus, for the case of a single real
boson we have two alternative identities (see eq.~(\ref{bosid}))
\be\label{3.21}
\prod_{n=1}^{\infty}\frac{1}{(1-q^n)} =
\Bigl(1+ \sum_{n=1}^{\infty}b^{({\rm wedge}) [\lambda=1]}_{([0^{n-1},1,0,\ldots,0];0)} (q) \Bigr) = 
\Bigl(1+ \sum_{n=1}^{\infty}b^{({\rm wedge}) [\lambda=0]}_{(0;[n,0,\ldots,0,n])} (q) \Bigr) \ . 
\ee
This fact will play an important role in the following. Since fermionisation and bosonisation involves
`non-perturbative' field-redefinitions, it is natural to express the bosonic and fermionic expansions
in terms of the representations $(\Lambda;0)$ and $(0;\Lambda)$, respectively, see
\cite{Gaberdiel:2012ku}. This motivates in retrospect the choice described in footnote~\ref{foot5}.

\subsection{Putting the Blocks Together}\label{sec:together}

Given  that we have identities for the partition functions of a complex boson and a complex fermion, 
see eqs.~(\ref{complxbos}) and (\ref{fersid2}), we can now easily obtain the expression for the
case with ${\cal N}=4$, see eq.~(\ref{Ndef}). Indeed, taking squares of the two expressions we conclude that 
\begin{eqnarray}
& & \prod_{n=1}^{\infty} \frac{(1 + y q^{n-\frac{1}{2}})^2 (1 + y^{-1} q^{n-\frac{1}{2}})^2}{(1-q^n)^4} \nonumber  \\[4pt]
& & \quad = \sum_{m_1,\ldots,m_4,n_1,\ldots,n_4} 
b^{({\rm wedge})[\lambda=1]}_{([0^{n_1-1},1,0,\ldots,0,1,0^{m_1-1}];0)}(q) \, \,
b^{({\rm wedge})[\lambda=1]}_{([0^{n_2-1},1,0,\ldots,0,1,0^{m_2-1}];0)}(q) \nonumber \\
& & \qquad \qquad \qquad \qquad \qquad b^{({\rm wedge})[\lambda=0]}_{(0;[n_3,0,\ldots,0,m_3])} (q,y) \, \,
b^{({\rm wedge})[\lambda=0]}_{(0;[n_4,0,\ldots,0,m_4])} (q,y) \ . \label{3.22}
\end{eqnarray}
Using the explicit expressions for these wedge characters, the product on the right-hand-side can be written as 
\be
\frac{q^{n_1+n_2+m_1+m_2+\frac{1}{2}( n_3^2+n_4^2+m_3^2+m_4^2)} \, y^{n_3+n_4-m_3-m_4}}
{\prod_{j=1}^{4} \prod_{k=1}^{n_j} (1-q^k) \, \prod_{k=1}^{m_j} (1-q^k) } \ . 
\ee
Eq.~(\ref{3.22}) now reproduces (\ref{wprim}) provided we identify 
\be\label{3.24}
\chi^{({\rm wedge}) [{\cal N}=4]}_{(0;[m,0,\ldots,0,0])} (q,y)  = 
\sum_{n_1+m_1+n_3+m_3=m} b^{({\rm wedge})[\lambda=1]}_{([0^{n_1-1},1,0,\ldots,0,1,0^{m_1-1}];0)}(q)  \,\, 
b^{({\rm wedge})[\lambda=0]}_{(0;[n_3,0,\ldots,0,m_3])} (q,y)  \ , 
\ee
and similarly for $\chi^{({\rm wedge}) [{\cal N}=4]}_{(0;[0,0,\ldots,0,n])} (q,y)$, as well as their products. It follows
from straightforward combinatorical considerations\footnote{We thank Constantin Candu for analysing this
in some detail.} that eq.~(\ref{3.24}) is indeed the wedge character of the ${\cal N}=4$ coset representation
$(0;[m,0,\ldots,0,0])$. This therefore effectively proves eq.~(\ref{wprim}).

\section{The Higher Spin Square}

In the previous section we have seen how the single particle generators of the stringy asymptotic symmetry algebra
can be organised in terms of representations of the wedge subalgebra ${\rm shs}_2[0]$
of the ${\cal N}=4$ superconformal ${\cal W}_\infty^{({\cal N}=4)}[0]$ algebra, 
see eq.~(\ref{3.22}) and below. These
single particle generators should now define a Lie algebra, the wedge algebra of the full stringy ${\cal W}$
algebra (i.e., the ${\cal W}$ algebra of the symmetric orbifold). This Lie algebra is the stringy
generalisation of the  ${\rm shs}_2[0]$ algebra that underlies the construction of the Vasiliev 
higher spin theory, and it should be thought of as describing the symmetries of string theory
on this background. 

The aim of this section is to make first steps towards characterising this global Lie algebra. 
Since we are viewing this question from the point of view of the dual CFT, the discussion will 
be mainly couched  in terms  of the full asymptotic symmetry algebra (whose wedge algebra
we are seeking to describe). We shall be able to give a fairly compelling and natural description of 
the full algebra in terms of a `higher spin square' for the the simpler symmetric orbifold of a single free boson, 
using the fact that we can equivalently describe the boson theory in terms of a complex fermion. We have also
found a neat finite-dimensional toy model in terms of the familiar Clifford algebra of 
gamma matrices, which exhibits nicely the exponential enhancement in rank.  However, while
it is easy to show that a similar structure also exists in the ${\cal N}=4$ case,
the constructions of the horizontal ${\cal W}_\infty$ algebra we have found so far are 
not particularly canonical.

\subsection{Free Bosons}\label{sec:free bosons}

Let us first begin with the simple case of the symmetric orbifold of a single free real boson, i.e., the theory of 
$N$ real bosons $\phi^j$ orbifolded by the permutation action $S_N$ on the colour index $j$. The symmetric
orbifold contains, in particular, the ${\rm SO}(N)$ invariant bilinears of the free boson generators, which
are known to generate ${\cal W}_{\infty}^{({\rm e})}[1]$. Furthermore, as we saw in eq.~(\ref{bosid1}), 
the remaining generators can be organised in terms of the representations $([0^{n-1},1,0,\ldots,0];0)$
of ${\cal W}_{\infty}^{({\rm e})}[1]$,
with $n=2$ corresponding to ${\cal W}_{\infty}^{({\rm e})}[1]$ itself. Note that it follows from 
eq.~(\ref{bosmodes}) that we can think of the generators coming from $([0^{n-1},1,0,\ldots,0];0)$
as describing the  symmetric invariants of $n$ powers of the field $\partial\phi^j$ 
with additional derivatives sprinkled on them (corresponding to the $k_i$ in eq.~(\ref{bosmodes})). 

In order to understand the resulting structure we now want to think of the 
generators of the full stringy ${\cal W}$ algebra as arranged in an 
infinite table or square as follows (see Figure~1).
We start with a horizontal row corresponding to the modes of $\sum_j (\partial \phi^j)^n$ with $n =1,2 \ldots$, labelling 
the columns, i.e., the representations $([0^{n-1},1,0,\ldots,0];0)$. The vertical column below each entry in this row 
consists of all the ${\cal W}_{\infty}^{({\rm e})}[1]$ descendants of this state, i.e., of all the states involving
$n$ bosons with an arbitrary number of derivatives sprinkled on them. As mentioned above, 
the $n=2$ column corresponds 
to the  ${\cal W}_{\infty}^{({\rm e})}[1]$ generators themselves. We will refer to this as the vertical higher spin or 
${\cal W}^{({\rm vert})}_{\infty}$ symmetry. 

What is intriguing about this description is that the top horizontal row of this square {\em also} 
generates a (distinct) ${\cal W}_{\infty}$ higher spin algebra --- which we will call the horizontal 
higher spin algebra ${\cal W}_{\infty}^{({\rm hor})}$.
In fact, the $\sum_j(\partial\phi^j)^n$ fields (corrected by terms 
involving a smaller number of powers of $(\partial\phi^j)$ and additional derivatives) are what one obtains by 
bosonising the fermionic bilinears (built from a complex fermion and its conjugate) that generate a 
${\cal W}_{1+\infty}$ algebra \cite{Pope:1991ig}. From this perspective, the 
$\sum_j(\partial\phi^j)^n$ terms correspond to the bilinear generators of the form
\be
\sum_l \bar{\psi}^l\, \partial^{n-1}\psi^l\ ,
\ee
or rather a sum of such bilinear terms with the $n-1$ derivatives distributed between $\psi^l$ and 
$\bar{\psi}^l$. The stress tensor corresponds to the $n=2$ term which is common to the vertical 
${\cal W}_{\infty}^{({\rm e})}[1]$ algebra but the terms with $n\neq 2$ are not part of the vertical algebra. 
Note that the correction terms mentioned above are actually necessary for the operators to be quasi-primary 
with respect to the stress tensor; for example, for the first few spins one finds explicitly \cite{Pope:1991ig}
\begin{eqnarray}\label{horspin}
S^{(1)} &= {\displaystyle \sum_j\partial\phi ^j\ , }\qquad  
& S^{(2)}  = \tfrac{1}{2} \sum_j (\partial\phi^j)^2\ , \\
S^{(3)} &=  {\displaystyle \tfrac{1}{3} \sum_j (\partial\phi^j)^3\ , } \quad  
& S^{(4)}  = \tfrac{1}{4} \sum_j (\partial\phi^j)^4 
-\tfrac{3}{20} \sum_j (\partial^2\phi^j)^2  
+ \tfrac{1}{10} \sum_j(\partial\phi^j)(\partial^3\phi^j) \ . \nonumber
\end{eqnarray}
Thus we conclude that the stringy ${\cal W}$-algebra contains actually {\em two} higher spin algebras:
the usual bilinear ${\cal W}_{\infty}^{(\rm e)}[1]$ algebra --- what we called the vertical algebra above --- and
the ${\cal W}_{1+\infty}$ algebra --- what we called the horizontal algebra above -- generated by the higher powers of $(\partial \phi)^n$, that contains
in particular ${\cal W}_{\infty}[0]$, see, e.g.,  \cite{Gaberdiel:2013jpa}. The two algebras
do not commute with one another, but rather generate the full algebra upon taking successive commutators,
i.e., what we have called the `higher spin square'. This structure is somewhat reminiscent of a Yangian symmetry 
that is believed to be present in 
${\cal N}=4$ super Yang-Mills theory at the planar level \cite{Drummond:2009fd}.

In principle, this formulation characterises all commutators of the stringy algebra recursively, and one can
also restrict this description to the wedge algebra (where instead of ${\cal W}_\infty^{({\rm  e})}[1]$ and
${\cal W}_\infty[0]$ we deal with the wedge subalgebras ${\rm hs}^{({\rm e})}[1]$ and ${\rm hs}[0]$,
respectively). However, unfortunately, we have not yet managed to find any
closed form expressions for these commutators. It would be very interesting to understand the structure of this algebra in more detail. 


We should emphasise that the above analysis is only correct in the large $N$ limit; for finite $N$
the relevant ${\cal W}$ algebras are truncated to finitely generated algebras (as follows, for example,
from the fact that their central charge is finite). However, unfortunately, these truncations are rather
complicated and in general not explicitly known.

%
%
%

\subsection{The Clifford Algebra Square -- a Toy Model}

In order to get a feeling for how this construction works explicitly, it may be useful to consider a 
finite toy model that also exhibits the same kind of structure. Consider the 
gamma matrix algebra in an even number of (euclidean) dimensions $2n$,
\be
\{\gamma^i, \gamma^j \}=2\delta^{ij} \ , \qquad (i,j =1,\ldots, 2n)\ .
\ee
Then, as is familiar, the terms with $k$ distinct gamma matrices $(1 \leq k \leq 2n)$ correspond to the 
$k^{\rm th}$ antisymmetric representation of $\mathfrak{so}(2n)$ and are of the form 
$\gamma^{[i_1}\gamma^{i_2}\cdots \gamma^{i_k]}$. Let us denote this subset of the full Clifford algebra 
as $H^{(k)}$, with $H^{(0)}$ denoting the identity. 
Since each $H^{(k)}$ is a representation of $\mathfrak{so}(2n)$, there is a highest weight state and a 
`vertical' column of $\mathfrak{so}(2n)$ descendants. 


In the context of the Clifford construction, there is a natural commutator one can define 
on the level of the generators in $H^{(k)}$. This can either be done abstractly via the so-called Clifford
commutators, see, e.g., \cite{LachiezeRey:2010au} for an introduction, or more explicitly since we can think of the 
$\gamma$-matrices as matrices of dimension $2^{n}$. On the level of vector spaces, the resulting commutators 
have the structure 
\be\label{cliffstru}
[H^{(k)}, H^{(m)}] = \sum_{r={\rm max}(0, (\frac{k+m}{2}-n))}^{{\rm min}(k,m)} 
\bigl(1- (-1)^{r+km} \bigr) \, H^{(k+m-2r)}\ .
\ee
This therefore gives a very concrete realisation of the commutation relations of these additional 
horizontal generators (as well as their associated columns). Note that the term with $k+m-2r=0$
never arises since in that case $r=m=k$, and $r + km$ is necessarily even.

The $\gamma$-matrices can be thought of as matrices in dimension $2^{n}$, and since 
\be
\dim \Bigl( \bigoplus_{k=0}^{2n}H^{(k)} \Bigr) = 2^{2n} \ ,
\ee
we deduce that the elements of the whole Clifford algebra generate the full space of these matrices.
Excluding the term with $k=0$ that never arises in commutators, we conclude that the full Lie 
algebra is $\mathfrak{su}(2^n)$.


We should mention that the resulting structure is again very 
similar to what we saw in the previous section. In particular, the rank of this Lie 
algebra grows exponentially with the number of species $n$, whereas the rank of the vertical 
Lie algebra $\mathfrak{so}(2n)$ is linear in $n$. Again, this is nicely parallel to what one should
expect for the relation between the higher spin symmetry and the stringy symmetry. 
The  number of higher spin symmetry generators have a linear growth in the number
of free fields, e.g., the theory based on $N$ free fermions leads to ${\cal W}_{1+N}$ 
with $N+1$ generating fields,
whereas the stringy symmetry generators have an exponential (Cardy) growth as one increases
$N$ in the symmetric product orbifold by $S_N$. 
While we have described the horizontal action of commutators between the various columns in 
eq.~(\ref{cliffstru}), we have not quite identified a small horizontal algebra (with rank of order $n$) 
analogous to ${\cal W}_{\infty}^{({\rm hor})}$. It would perhaps be interesting to do so.  
\smallskip

\subsection{The ${\cal N}=4$ Higher Spin Square}\label{sec:susyhspsq}

For the case with ${\cal N}=4$ we should expect, similar to the bosonic case, 
to be able to construct
a `horizontal' ${\cal W}_\infty$ algebra that will generate, together with the
vertical ${\cal W}_{\infty}^{(4)}[0]$ algebra, the full set of generators (\ref{wedgegen})
of the stringy extension. In particular, the horizontal algebra should contain at least
one generator from each of the representations $(0;[m,0,\ldots,0,n])$ with $m,n\geq 0$. 

One possible choice for such a horizontal algebra is to consider the 
algebra that is generated by the various powers of $(\partial \phi^1)$ and
$(\partial\bar\phi^1)$, where $(\phi^1,\bar{\phi}^1)$ is one arbitrarily chosen complex boson.
The generator $(\partial \phi^1)^m \, (\partial\bar\phi^1)^n$ will then contribute to the 
representation $(0;[m,0,\ldots,0,n])$, and hence this construction will account for all the generators.
Furthermore, up to lower order correction terms, these generators define
a bosonic ${\cal W}_\infty$ algebra, as follows from the same arguments as in 
Section~\ref{sec:free bosons}. It is also clear from the analysis of Section~\ref{sec:together} that
the full generating function can be expanded in terms of the corresponding characters, using 
twice the identity eq.~(\ref{3.21}).

However, there are a few rather unnatural aspects about this construction. First of all,
the horizontal algebra is not really of higher spin type since the number of generators of 
spin $s$ grows linearly with $s$ --- the relevant generators are schematically of the form
\be
(\partial \phi^1)^m \, (\partial\bar\phi^1)^{s-m} \ , \qquad m=0,\ldots, s \ .
\ee 
One way to avoid this, is to make the vertical algebra slightly larger by
including all bilinear terms of the underlying free fields, i.e., by adding in the 
representations $(0;[2,0,\ldots,0])$ and $(0;[0,\ldots,0,2])$. (It is fairly obvious
that this will also lead to a consistent ${\cal W}_\infty$ algebra whose number of
generators does not grow with the spin.) Then we can make the horizontal algebra
slightly smaller by only including the powers of a single real boson $(\partial \Phi^1)^m$.
The resulting horizontal higher spin algebra is then the same as the one analysed in 
Section~\ref{sec:free bosons}, i.e., it defines the bosonic algebra ${\cal W}^{({\rm e})}_\infty[1]$. 
In particular, it is therefore of higher spin type. 

However, both of these constructions are somewhat unnatural since the resulting
horizontal algebras are purely bosonic, and in particular, do not
contain the ${\cal N}=4$ superconformal algebra. It is possible to overcome this problem,
but the only way we have found so far, is again rather ad hoc. Instead of the above algebras 
we consider the ${\cal W}_\infty$ algebra that is generated by bilinears of the generators 
\be
(\partial \Phi^1)^{m_1} \ , \qquad \quad (\partial^{m_i} \Phi^i) \ , \ \  i = 2,3,4 \ , \qquad \qquad 
(\partial^{r_j} \psi^j) \ , \ \ j=1, 2, 3, 4 \ ,
\ee
where the $\Phi^j$, with $j=1,2,3,4$, denote the $4$ real bosonic fields, and the $\psi^j$ are the corresponding
fermionic partners. The resulting algebra contains the ${\cal N}=4$ superconformal
algebra, and its structure is fairly similar to that of the bilinear extension of ${\cal W}^{(4)}_\infty[0]$ --- in particular, 
the number of generators does not grow with spin, and the algebra is of `higher spin' type. 
Again, it is clear that the  full generating function can be expanded in terms of the corresponding 
characters, using  as before  the identity eq.~(\ref{3.21}). However, the construction singles
out one of the four  real boson fields, which is not particularly natural. So while these 
considerations demonstrate
that also in the ${\cal N}=4$ case, the higher spin square can be generated by two higher
spin algebras, we suspect that there exists a more natural (and symmetrical) construction.


\section{Concluding Remarks}

In this paper we have analysed the 
symmetry algebra of string theory on ${\rm AdS}_3\times {\rm S}^3\times \mathbb{T}^4$ in the tensionless
limit. We have found that the unbroken symmetries in this ${\rm AdS}_3$ vacuum exhibit the structure 
of a higher spin square which we have described. The lesson seems to be (at least as far as these symmetries 
are concerned) that knowing the commutators of the horizontal and vertical higher spin algebras completely 
determine all commutators of the exponentially larger unbroken stringy symmetry algebra. This may indicate that 
higher spin symmetries play an important role in string theory, beyond just 
describing the dynamics of the leading Regge trajectory. 

Of course, one would hope to be able to exploit the power of this symmetry to deduce more about the 
broken symmetry or higgsed phase. 
For this it will be important to go beyond the in-principle determination of 
commutators mentioned above to more explicit characterisations of the algebra. It would also be
interesting to understand the relation of this symmetry algebra to the Yangian that is believed to 
control the planar limit of ${\cal N}=4$ super Yang Mills, as well as to the spin chain analysis of 
\cite{Borsato:2014exa,Sax:2014mea}, see also \cite{Sfondrini:2014via} for a recent review. 
It would be good to see whether the multi-particle algebras, studied recently by Vasiliev \cite{Vasiliev:2012tv} as a way to generalise higher spin symmetry in string theory, have a connection with the higher spin square structure. 
A related question is 
to understand the representations of the higher spin square, especially those relevant for the description of the 
fields of string theory. In \cite{Gaberdiel:2014cha} we described how the twisted 
and untwisted sectors of the symmetric orbifold theory can be organised in terms of representations of 
what we would now call the vertical  ${\cal W}_\infty$ algebra. Once again, after considering the 
single particle excitations, these must assemble into bigger representations of the stringy symmetry 
above. It would be interesting to see how many such representations appear. 
One should also be able to use this information to constrain matrix elements of 
the perturbation by the $\mathfrak{su}(2)_R$ singlet twist two operator 
{\it a la} Wigner-Eckart. A somewhat different approach to studying such perturbations might be to see whether current algebra methods, so successful for vector like theories \cite{Maldacena:2012sf}, can be adapted to the case with a larger stringy symmetry and give useful constraints.  

Another direction, which we have only touched on, concerns the finite truncation of the higher spin square. As mentioned, 
the ${\cal W}$-algebras at finite $N$ will be finitely generated and there are relations between hitherto 
(at $N=\infty$) independent generators. This truncation is not entirely straightforward even for a symmetric 
orbifold of $N$ free bosons. Such a deformed algebra reflects a quantum modification of the symmetry algebra 
and is presumably responsible for non-perturbative effects like the stringy exclusion principle. 
It would be interesting to see if the Clifford algebra square arises naturally in any such truncation. 
Another 
modification comes into play when we consider the theory on a finite $\mathbb{T}^4$. We have been 
implicitly considering the theory on $\mathbb{R}^4$ by neglecting modes which carry charge on the 
$\mathbb{T}^4$ ---  as usually done in the D1-D5 system. Putting the theory on $\mathbb{T}^4$ again 
breaks part of the symmetry, and one may study how the corresponding perturbation is constrained by the 
unbroken symmetry. 

The unbroken symmetries described here are those of a particular background of string theory. But the 
underlying symmetries of the theory are independent of the choice of background --- the background enters 
only in manifesting which symmetries are unbroken. One may ask whether the higher spin square described 
here is the largest possible unbroken symmetry. As mentioned in the introduction, in higher dimensional 
${\rm AdS}$ vacua, the maximal unbroken symmetries are smaller in size (though not subalgebras of the 
above stringy algebra). Are there backgrounds which manifest 
even larger unbroken symmetries? Even amongst ${\rm AdS}_3$ vacua there are others (such as 
the case where $\mathbb{T}^4$ is replaced by K3) which will have different unbroken stringy symmetries at, 
say, the orbifold point. Finally, necessary and sufficient conditions for 2d permutation orbifold CFTs to have a 
holographic dual have recently been formulated in \cite{Haehl:2014yla,Belin:2014fna} (using results of 
\cite{Hartman:2014oaa}). It would be nice to extend the above  results for the symmetric orbifold to the more 
general permutation orbifolds with potential holographic duals. 

\section*{Acknowledgements}

We thank Niklas Beisert, Constantin Candu, Avinash Dhar, Shiraz Minwalla, Suvrat Raju, Ashoke Sen, Zhenya Skvortsov, 
Spenta Wadia, and Misha Vasiliev
for helpful discussions. The work of M.R.G.\ is partially supported by the Swiss National Science Foundation, 
in particular through the NCCR SwissMAP, and he thanks Cambridge University and the IAS in Princeton for 
hospitality during various stages of this work. 
R.G., as always, expresses his gratitude to the people of India 
for supporting the enterprise of theoretical physics. He also thanks the Pauli Center of 
ETH Zurich and ICTS-TIFR, Bangalore for hospitality during different stages of this project. 

%

\bibliographystyle{JHEP}

\begin{thebibliography}{99}


\bibitem{Gross:1988ue} 
D.J.~Gross,
``High-energy symmetries of string theory,''
Phys.\ Rev.\ Lett.\  {\bf 60} (1988) 1229.
  
\bibitem{Witten:1988zd} 
E.~Witten,
``Space-time and topological orbifolds,''
Phys.\ Rev.\ Lett.\  {\bf 61} (1988) 670.
  
\bibitem{Moore:1993qe} 
G.W.~Moore,
``Symmetries and symmetry breaking in string theory,''
in proceedings of the SUSY '93 conference, `Supersymmetry and unification of fundamental interactions,'
(1993) 540 {\tt [arXiv:hep-th/9308052]}.
 
 \bibitem{Sagnotti:2011qp} 
A.~Sagnotti,
``Notes on strings and higher spins,''
J.\ Phys.\ A {\bf 46} (2013) 214006 
{\tt [arXiv:1112.4285 [hep-th]]}.

\bibitem{Sundborg:2000wp}
B.~Sundborg,
``Stringy gravity, interacting tensionless strings and massless higher spins,"
Nucl.\ Phys.\ Proc.\ Suppl.\  {\bf 102} (2001) 113
{\tt [arXiv:hep-th/0103247]}.
  
\bibitem{Witten}
E.~Witten, talk at the John Schwarz 60-th birthday symposium (Nov. 2001), \newline
{\tt http://theory.caltech.edu/jhs60/witten/1.html}.

\bibitem{Mikhailov:2002bp}
A.~Mikhailov,
``Notes on higher spin symmetries,''
{\tt arXiv:hep-th/0201019}.

\bibitem{Sezgin:2002rt}
E.~Sezgin and P.~Sundell,
``Massless higher spins and holography,"
Nucl.\ Phys.\  B {\bf 644} (2002) 303
[Erratum-ibid.\  B {\bf 660} (2003) 403]
{\tt [arXiv:hep-th/0205131]}.

\bibitem{Vasiliev:2003ev}
M.A.~Vasiliev,
``Nonlinear equations for symmetric massless higher spin fields in (A)dS(d),"
Phys.\ Lett.\  B {\bf 567} (2003)  139
{\tt [arXiv:hep-th/0304049]}.

 \bibitem{Klebanov:2002ja}
 I.R.~Klebanov and A.M.~Polyakov,
``AdS dual of the critical O(N) vector model,"
Phys.\ Lett.\  B {\bf 550} (2002) 213
{\tt [arXiv:hep-th/0210114]}.

\bibitem{Sezgin:2003pt} 
E.~Sezgin and P.~Sundell,
``Holography in 4D (super) higher spin theories and a test via cubic scalar couplings,''
JHEP {\bf 0507} (2005) 044
{\tt  [arXiv:hep-th/0305040]}.

\bibitem{Giombi:2009wh}
S.~Giombi and X.~Yin,
``Higher spin gauge theory and holography: the three-point functions,"
JHEP {\bf 1009} (2010) 115 {\tt [arXiv:0912.3462 [hep-th]]}.

\bibitem{Giombi:2010vg}
S.~Giombi and X.~Yin,
``Higher spins in AdS and twistorial holography,"
 JHEP {\bf 1104} (2011) 086 {\tt [arXiv:1004.3736 [hep-th]]}.

\bibitem{Gaberdiel:2010pz}
M.R.~Gaberdiel and R.~Gopakumar,
 ``An AdS$_3$ dual for minimal model CFTs,''
Phys.\ Rev.\ D {\bf 83} (2011) 066007
 {\tt [arXiv:1011.2986 [hep-th]]}.
 
\bibitem{Aharony:2011jz}
O.~Aharony, G.~Gur-Ari and R.~Yacoby,
``d=3 bosonic vector models coupled to Chern-Simons gauge theories,''
JHEP {\bf 1203} (2012) 037
{\tt [arXiv:1110.4382 [hep-th]]}.

\bibitem{Giombi:2011kc} 
 S.~Giombi, S.~Minwalla, S.~Prakash, S.P.~Trivedi, S.R.~Wadia and X.~Yin,
 ``Chern-Simons theory with vector fermion matter,''
Eur.\ Phys.\ J.\ C {\bf 72} (2012) 2112 
 {\tt [arXiv:1110.4386 [hep-th]]}.
 
\bibitem{Giombi:2013fka} 
S.~Giombi and I.R.~Klebanov,
``One loop tests of higher spin AdS/CFT,''
 JHEP {\bf 1312} (2013) 068 
{\tt [arXiv:1308.2337 [hep-th]]}.
  
\bibitem{Giombi:2014iua} 
S.~Giombi, I.R.~Klebanov and B.R.~Safdi,
``Higher spin AdS$_{d+1}$/CFT$_d$ at one loop,''
Phys.\ Rev.\ D {\bf 89} (2014) 084004 
{\tt [arXiv:1401.0825 [hep-th]]}.

\bibitem{Gaberdiel:2012uj}
M.R.~Gaberdiel and R.~Gopakumar,
``Minimal model holography,''
J.\ Phys.\ A: Math.\ Theor.\ {\bf 46} (2013) 214002
{\tt [arXiv:1207.6697 [hep-th]]}.

\bibitem{Giombi:2012ms} 
S.~Giombi and X.~Yin,
``The higher spin/vector model duality,''
J.\ Phys.\ A {\bf 46} (2013) 214003 
{\tt [arXiv:1208.4036 [hep-th]]}.

\bibitem{David:2002wn} 
 J.R.~David, G.~Mandal and S.R.~Wadia,
``Microscopic formulation of black holes in string theory,''
Phys.\ Rept.\  {\bf 369} (2002) 549 
{\tt  [arXiv:hep-th/0203048]}.  

\bibitem{Gaberdiel:2014cha} 
M.R.~Gaberdiel and R.~Gopakumar,
``Higher spins \& strings,''
JHEP {\bf 1411} (2014) 044 
{\tt [arXiv:1406.6103 [hep-th]]}.
  
\bibitem{Brown:1986nw}
J.D.~Brown and M.~Henneaux,
``Central charges in the canonical realization of asymptotic symmetries: an
example from three-dimensional gravity,"
Commun.\ Math.\ Phys.\  {\bf 104} (1986) 207.
  
\bibitem{Henneaux:2010xg}
M.~Henneaux and S.-J.~Rey,
``Nonlinear W(infinity) algebra as asymptotic symmetry of three-dimensional
higher spin Anti-de Sitter gravity,"
JHEP {\bf 1012} (2010) 007 
{\tt [arXiv:1008.4579 [hep-th]]}.

\bibitem{Campoleoni:2010zq}
A.~Campoleoni, S.~Fredenhagen, S.~Pfenninger and S.~Theisen,
``Asymptotic symmetries of three-dimensional gravity coupled to higher-spin fields,"
JHEP {\bf 1011} (2010) 007
{\tt  [arXiv:1008.4744 [hep-th]]}.

\bibitem{Gaberdiel:2011wb}
M.R.~Gaberdiel and T.~Hartman,
``Symmetries of holographic minimal models,''
JHEP {\bf 1105} (2011) 031 
{\tt [arXiv:1101.2910 [hep-th]]}.  

\bibitem{Campoleoni:2011hg}
A.~Campoleoni, S.~Fredenhagen and S.~Pfenninger,
``Asymptotic W-symmetries in three-dimensional higher-spin gauge theories,''
JHEP {\bf 1109} (2011) 113
{\tt  [arXiv:1107.0290 [hep-th]]}.
  
\bibitem{Gaberdiel:2013vva}
M.R.~Gaberdiel and R.~Gopakumar,
``Large $\mathcal{N}=4$ holography,''
 JHEP {\bf 1309} (2013) 036
{\tt  [arXiv:1305.4181 [hep-th]]}.

\bibitem{Pope:1991ig}
C.N.~Pope,
``Lectures on W algebras and W gravity,''
{\tt arXiv:hep-th/9112076}.

\bibitem{Dijkgraaf:1996xw}
R.~Dijkgraaf, G.W.~Moore, E.P.~Verlinde and H.L.~Verlinde,
``Elliptic genera of symmetric products and second quantized strings,''
Commun.\ Math.\ Phys.\  {\bf 185} (1997) 197
{\tt  [arXiv:hep-th/9608096]}.

\bibitem{Gaberdiel:2011nt}
M.R.~Gaberdiel and C.~Vollenweider,
``Minimal model holography for SO($2N$),''
JHEP {\bf 1108} (2011) 104
{\tt [arXiv:1106.2634 [hep-th]]}.

\bibitem{CGKV}
C.~Candu, M.R.~Gaberdiel, M.~Kelm and C.~Vollenweider,
``Even spin minimal model holography,''
JHEP  {\bf 1301} (2013) 185
 {\tt [arXiv:1211.3113 [hep-th]]}. 
 
\bibitem{Gaberdiel:2011zw}
M.R.~Gaberdiel, R.~Gopakumar, T.~Hartman and S.~Raju, 
``Partition functions of holographic minimal models,''
JHEP {\bf 1108} (2011) 077
{\tt [arXiv:1106.1897 [hep-th]]}.

\bibitem{AAR}
G.E.~Andrews, R.~Askey and R.~Roy,
``Special functions,"  
Cambridge University Press (1999).  


\bibitem{Gaberdiel:2011aa}
M.R.~Gaberdiel and P.~Suchanek,
``Limits of minimal models and continuous orbifolds,''
JHEP {\bf 1203} (2012) 104
{\tt [arXiv:1112.1708 [hep-th]]}.
 
\bibitem{Gaberdiel:2012ku}
M.R.~Gaberdiel and R.~Gopakumar,
``Triality in minimal model holography,''
JHEP {\bf 1207} (2012) 127 
{\tt [arXiv:1205.2472 [hep-th]]}.
 

\bibitem{Gaberdiel:2013jpa}
M.R.~Gaberdiel, K.~Jin and W.~Li,
``Perturbations of ${\cal W}_\infty$ CFTs,''
JHEP {\bf 1310} (2013) 162
{\tt [arXiv:1307.4087 [hep-th]]}.


\bibitem{Drummond:2009fd} 
J.M.~Drummond, J.M.~Henn and J.~Plefka,
``Yangian symmetry of scattering amplitudes in N=4 super Yang-Mills theory,''
JHEP {\bf 0905} (2009) 046
{\tt [arXiv:0902.2987 [hep-th]]}.

\bibitem{LachiezeRey:2010au}
M.~Lachieze-Rey,
``Spin and Clifford algebras, an introduction,''
{\tt arXiv:1007.2481 [gr-qc]}.

\bibitem{Borsato:2014exa}
R.~Borsato, O.~Ohlsson Sax, A.~Sfondrini and B.~Stefanski,
 ``Towards the all-loop worldsheet S matrix for AdS$_3\times {\rm S}^3\times \mathbb{T}^4$,''
Phys.\ Rev.\ Lett.\  {\bf 113} (2014) 131601
{\tt [arXiv:1403.4543 [hep-th]]}; 
%
``The complete AdS$_{3} \times$ S$^3 \times$ T$^4$ worldsheet S matrix,''
JHEP {\bf 1410} (2014) 66
{\tt [arXiv:1406.0453 [hep-th]]}.

\bibitem{Sax:2014mea}
O.~Ohlsson Sax, A.~Sfondrini and B.~Stefanski,
``Integrability and the conformal field theory of the Higgs branch,''
{\tt arXiv:1411.3676 [hep-th]}.

\bibitem{Sfondrini:2014via}
A.~Sfondrini,
``Towards integrability for AdS3/CFT2,''
J.\ Phys.\ A: Math.\ Theor.\ {\bf 48} (2015) 023001
{\tt [arXiv:1406.2971 [hep-th]]}.

\bibitem{Vasiliev:2012tv} 
 M.A.~Vasiliev,
``Multiparticle extension of the higher-spin algebra,''
Class.\ Quant.\ Grav.\  {\bf 30} (2013) 104006 
{\tt   [arXiv:1212.6071 [hep-th]]}.

\bibitem{Maldacena:2012sf} 
J.~Maldacena and A.~Zhiboedov,
``Constraining conformal field theories with a slightly broken higher spin symmetry,''
Class.\ Quant.\ Grav.\  {\bf 30} (2013) 104003 
{\tt [arXiv:1204.3882 [hep-th]]}.


\bibitem{Haehl:2014yla} 
F.M.~Haehl and M.~Rangamani,
``Permutation orbifolds and holography,''
{\tt  arXiv:1412.2759 [hep-th]}.
  
\bibitem{Belin:2014fna}
A.~Belin, C.A.~Keller and A.~Maloney,
``String universality for permutation orbifolds,''
{\tt arXiv:1412.7159 [hep-th]}.
    
\bibitem{Hartman:2014oaa} 
T.~Hartman, C.A.~Keller and B.~Stoica,
``Universal spectrum of 2d conformal field theory in the large c limit,''
JHEP {\bf 1409} (2014) 118 
{\tt [arXiv:1405.5137 [hep-th]]}.
  
\end{thebibliography}

\end{document}